\documentclass[prd,preprint,aps,eqsecnum,nofootinbib,showpacs,floatfix]{revtex4-1}

\usepackage{amsmath}    
\usepackage{amssymb}
\usepackage{bm}         
\usepackage{graphicx}   
\usepackage{verbatim}   
\usepackage{color}      
\usepackage{subfigure}  
\usepackage{dsfont}
\usepackage{multirow}

\definecolor{orange}{rgb}{1.0, 0.647059, 0.0}
\definecolor{newmagenta}{rgb}{1.0, 0.0, 1.0}
\definecolor{darkgreen}{rgb}{0.0, 0.545098, 0.0}

\newcommand{\Dslash}{\ensuremath{D\kern-0.65em/\kern0.15em}}
\newcommand{\Eslash}{\ensuremath{E\kern-0.65em/\kern0.15em}}
\newcommand{\dslash}{\ensuremath{\partial\kern-0.65em/\kern0.15em}}
\newcommand{\vslash}{\ensuremath{v\kern-0.65em/\kern0.15em}}
\newcommand{\eslash}{\ensuremath{\kern0.1em\epsilon\kern-0.4em/\kern-0.1em}}

\newcommand{\bra}[1]{\langle #1 |}
\newcommand{\ket}[1]{| #1 \rangle}
\newcommand{\Tr}{\mathop{\mathrm{Tr}}}
\newcommand{\diag}{\mathop{\text{diag}}\nolimits}

\newcommand{\overbar}[1]{\mkern 1.5mu\overline{\mkern-1.5mu#1\mkern-1.5mu}\mkern 1.5mu}

\begin{document}
\title{Variational method with staggered fermions}
\author{Carleton DeTar and Song-Haeng Lee}
\affiliation{Department of Physics and Astronomy, University of Utah, Salt Lake City, Utah}

\begin{abstract}
The variational method is used widely for determining eigenstates of
the QCD hamiltonian for actions with a conventional transfer matrix,
{\it e.g.}, actions with improved Wilson fermions.  An alternative
lattice fermion formalism, staggered fermions, does not have a
conventional single-time-step transfer matrix.  Nonetheless, with a
simple modification, the variational method can also be applied to
that formalism.  In some cases the method also provides a mechanism
for separating the commonly paired parity-partner states.  We discuss
the extension to staggered fermions and illustrate it by applying it
to the calculation of the spectrum of charmed-antistrange mesons
consisting of a clover charm quark and a staggered strange antiquark.
\end{abstract}

\maketitle

\section{Conventional Wilson Variational Method}

We consider the hadronic correlator for propagation from Euclidean
time 0 to time $t$ (an integer in lattice units):
\begin{equation}
   C_{ij}(t) = \bra{0}{\cal O}_i(t) {\cal O}^\dagger_j(0) \ket{0} \,,
\end{equation}
generated by a set of hermitian interpolating operators ${\cal
  O}_i(t)$ and propagating according to the QCD hamiltonian $H$
derived from an action with a single time-step transfer matrix $T =
\exp(-H)$.  We assume the time extent of the lattice is sufficiently
large that we may consider propagation only forward in time.  Then,
eigenstates of the hamiltonian with energy $E_n$ correspond to
eigenstates of the transfer matrix $T$ with eigenvalue
$\exp(-E_n)$. (We enumerate energies in ascending order.)  In terms of
these eigenstates, the correlator has a multiexponential eigenenergy
representation
\begin{equation}
   C_{ij}(t) = \sum_n \bra{0}{\cal O}_i(t)\ket{n} e^{-E_n t} \bra{n} {\cal O}^\dagger_j(0) \ket{0} \,,
\label{eq:multiexp}
\end{equation}
or in matrix form
\begin{equation}
   C(t) = Z T^t Z^\dagger \,,
\end{equation}
where the overlap matrix is 
\begin{equation}
  Z_{i,n} = \bra{0}{\cal O}_i(t)\ket{n} \,.
\end{equation}
In a typical application $C(t)$ is known and we want to determine the
energies $E_n$.  We start by truncating the infinite sum in
Eq.~(\ref{eq:multiexp}) to a finite sum for $n \in [1,N]$ and
introduce at least $N$ linearly independent interpolating operators
${\cal O}_i(t)$.  Then we can find the energies by solving the
generalized eigenvalue problem
\begin{equation}
  C(t)u_n = \lambda_n(t,t_0) C(t_0)u_n \,,
\end{equation}
where $u_n$ is the $k$th column of the matrix $Z$, and
$\lambda_n(t,t_0)$ is an approximation to the eigenvalue
$\exp[-E_n(t - t_0)]$ of the infinite transfer matrix $T^{t-t_0}$.
They are approximations, because truncating the multiexponential sum
introduces errors \cite{Luscher:1990ck,Blossier:2009kd}.  We discuss
the approximation at greater length below.

\section{Staggered Variational Method}
\label{sec:variational}

When the hadronic correlator involves staggered fermions, the
multiexponential expansion includes terms that oscillate in
time:
\begin{equation}
  C_{ij}(t) = \sum_n \bra{0}{\cal O}_i(t)\ket{n} s_n(t) e^{-E_n t} 
    \bra{n} {\cal O}^\dagger_j(0) \ket{0} \,.
\label{eq:staggmultiexp}
\end{equation}
where the $t$-dependent sign $s_n(t) = 1$ for a nonoscillating state
$n$ and $s_n(t) = (-1)^t$ for an oscillating state.

This oscillation is well known for mesons and baryons constructed from
single-time-slice interpolating operators consisting of only staggered
fermions \cite{Golterman:1986jf,Golterman:1984dn}.  The oscillating
component corresponds to a state with parity opposite to that of the
nonoscillating component.  Since the states often come in pairs, they
are sometimes called ``parity partners''.  In the case of a meson with
definite charge conjugation, the partner also has the opposite charge
conjugation quantum number.

We are interested here in the correlator for a meson arising from a
source interpolating operator consisting of a Dirac (Wilson or clover)
quark and a staggered antiquark.  To construct the hadronic
correlator, we first convert the staggered propagator $S(x^\prime, x)$
to a ``naive'' propagator \cite{Wingate:2002fh}, using
\begin{equation}
  N(x^\prime, x) = \Gamma^\dagger(x^\prime) \Gamma(x) S(x^\prime, x) \,,
\label{eq:naive}
\end{equation}
where, in one convention,
\begin{equation}
  \Gamma(x) = \gamma^{x_1} \gamma^{x_2} \gamma^{x_3} \gamma^{x_0} \, .
\end{equation}
It is now standard practice to work with improved staggered fermion
propagators $S$ so the resulting ``naive'' propagator $N$ inherits the
improvement.

The resulting propagator $N(x^\prime, x)$ carries both color and spin
indices and so can be treated on the same footing as the propagator
for the Dirac quark $W(y^\prime, y)$.  So, for example, if the source
interpolating operator is a local zero-momentum quark-antiquark
bilinear with gamma matrix $\Gamma_A$, and, similarly, the sink
interpolating operator is a local bilinear with gamma matrix
$\Gamma_B$, then the resulting hadronic correlator with $x^\prime =
(t, {\bf x^\prime})$ and $x = (0, {\bf x})$ is
\begin{equation}
  C(t) = \sum_{\bf x} \Tr[\Gamma_B N(x^\prime, x) \Gamma_A^\dagger W(x, x^\prime)] \,,
\end{equation}
where the trace is over both spins and colors. 

Now consider a corresponding correlator $C^\prime(t)$ with the source
and sink gamma matrices replaced with $\Gamma_A \gamma_0 \gamma_5$ and
$\Gamma_B \gamma_0 \gamma_5$, respectively.  This replacement
preserves the angular momentum, but reverses the parity of the state
and its charge conjugation quantum number, if relevant.  It is easy to
show that
\begin{equation}
  C^\prime(t) = C(t)(-)^t \,\, .
\end{equation}
because 
\begin{equation}
  \gamma_0 \gamma_5 N(x^\prime, x) = (-)^{x_0^\prime - x_0} N(x^\prime, x)\gamma_0\gamma_5 \,.
\end{equation}
Thus with meson correlators involving staggered fermions, there is a
symmetry relating correlators for channels of opposite $C$ and $P$
quantum numbers.  A given state appears in both correlators, in one of
them with no oscillation and in the other, with oscillation.

With the single-time-slice Dirac-plus-staggered interpolating
operators we have studied, hadronic correlators typically contain both
oscillating and nonoscillating contributions as contemplated in
Eq.~(\ref{eq:staggmultiexp}). From the discussion of the meson case
above, we see that oscillating contribution is associated with a
partner state of the opposite $P$ and $C$. Moreover, if an
interpolating operator ${\cal O}_i$ is constructed from a hermitian
bilinear with gamma matrix $\Gamma_A$, the operator constructed from
$\Gamma_A \gamma_0 \gamma_5$ is antihermitian.
Thus
\begin{equation}
  \bra{n} {\cal O}_i(0) \ket{0} = -\bra{n} {\cal O}^\dagger_i(0) \ket{0} \, .
\end{equation}
A consequence is that the parity partner contributions, in addition to
oscillating with a factor $(-)^t$ include an overall minus sign from
the antihermiticity noted above.  Thus, the correlator has the matrix
form
\begin{equation}
    C(t) = Z T^t g Z^\dagger \,,
\end{equation}
where $T = g \diag{e^{-E_n}}$ and $g = \diag{s_n(1)}$, that is,
a diagonal matrix with a plus (minus) sign for nonoscillating
(oscillating) states.

The generalized eigenvalue problem is the same as before:
\begin{equation}
  C(t)u_n = \lambda_n(t,t_0) C(t_0) u_n  \, ,
\end{equation}
but with oscillating as well as nonoscillating eigenvalues
$\lambda_n(t,t_0)$.  We modify the ordering convention so the
eigenvalues are in decreasing order according to their magnitudes
$|\lambda_n(t,t_0)| > |\lambda_{n+1}(t,t_0)|$ for large $t$ and $t_0$.
If there are $N$ linearly independent interpolating operators and the
multiexponential expansion terminates at the the $N$th energy, the
generalized eigenvalue problem yields $\lambda_n = s_n(t-t_0) e^{-E_n
  (t-t_0)}$ exactly.  Of course, in practice, the multiexponential
expansion does not terminate, so the generalized eigenvalues only
approximate $s_n(t-t_0)e^{-E_n (t-t_0)}$.  The ALPHA Collaboration
used perturbation theory to treat the effect of restoring energy
levels with $E_n > E_N$ \cite{Blossier:2009kd}.  Their analysis is
easily generalized to the present case with oscillating and
nonoscillating states.  To second order we have
\begin{eqnarray}
  \lambda_n(t,t_0) & \approx & s_n(t-t_0) (1-a_n(t_0)) e^{-E_n (t - t_0)} \nonumber\\
   & + & \sum_{m>n}^N b_{m,n}(t_0) s_m(t-t_0) e^{-E_m (t - t_0)} \\
   & - & \sum_{m<n} b_{m,n}(t_0) s_m(t-t_0) e^{-(2 E_n - E_m)(t-t_0)} +  {\cal O}(e^{- E_{N+1}(t-t_0)}) \,.
   \nonumber
  \label{eq:eigenvalue}
\end{eqnarray}
The coefficients $a_n$ and $b_{m,n}$ depend only upon 
$t_0$ and overlap factors:
\begin{eqnarray}
  a_n(t_0) & \approx & A_{n,n,N+1}s_n(t_0)s_{N+1}(t_0)e^{-(E_{N+1}-E_n)t_0} \nonumber \\
 &-& \left[ e^{-2(E_{N+1}-E_n)t_0}|A_{n,n,N+1}|^2 + \sum_{m > n}^N b_{n,m,N+1}(t_0)\right]
	\label{eq:eigenvalueamp01} \\
  b_{m,n}(t_0) & \approx &  |A_{m,n,N+1}|^2 s_n(t_0)s_m(t_0)e^{-(2 E_{N+1} - E_n - E_m)t_0} \,,
  \label{eq:eigenvalueamp02}
\end{eqnarray}
where $A_{m,n,N+1}$ is given by the product of overlaps
\begin{equation}
  A_{m,n,N+1} = \left(\sum_{i=1}^N u_{m,i}^* Z_{i,N+1}\right)
       \left(\sum_{i=1}^N Z_{i,N+1}^* u_{n,i}\right) \, .
  \label{eq:overlapratio}
\end{equation}

As the number $N$ of linearly independent interpolating operators is
increased at fixed $t$, $t_0$, $m$, and $n$, the factors
$e^{-(E_{N+1}-E_n)t_0}$ decrease exponentially, so the coefficients
$a_n$ and $b_{m,n}$ vanish exponentially.  So, as expected,
\begin{equation}
  \lambda_n(t,t_0) \rightarrow s_n(t-t_0) e^{-E_n (t - t_0)} \, .
\end{equation}
Alternatively, if $t_0$ is large for fixed $N$, $m$, and $n$, the
exponential factors also suppress the coefficients $a_n$ and $b_{m,n}$
with the same result.  In Ref.~\cite{Blossier:2009kd} the ALPHA
collaboration argued that to assure a plateau in the ``effective
energy'', {\it i.e.}, to obtain a good approximation to the above
asymptotic form, one should require $t_0 > t-t_0 \gg 0$.  However,
making $N$ large increases the cost of the calculation, and it is not
always possible to make $t_0$ large and still have a good signal for
the hadron correlator.  For this reason the Hadron Spectrum
Collaboration, in a more conventional application without staggered
fermions, proposed fitting eigenvalues to the form
\begin{equation}
  \lambda_n(t,t_0) \approx (1-a_n) e^{-E_n (t - t_0)} + a_n e^{-\bar E_n (t - t_0)} \,,
\end{equation}
where the second term approximates higher corrections
\cite{Liu:2012ze}.  With staggered fermions, we simply include terms
that oscillate in $t$, as, for example with the model
\begin{eqnarray}
  \lambda_n(t,t_0) &\approx& [1-a_n(t_0)] s_n(t-t_0) e^{-E_n (t - t_0)} + 
     b_n(t_0) s_n(t-t_0)e^{-\bar E_n (t - t_0)} + \nonumber \\
     &+& c_n(t_0) s_n^\prime(t-t_0) e^{-E_n^\prime (t - t_0)} 
    + d_n(t_0) s_n^\prime(t-t_0) e^{-\bar E_n^\prime (t - t_0)} \,,
\label{eq:fitstaggeigen}
\end{eqnarray}
where $s_n^\prime(t)$ oscillates if $s_n(t)$ does not, and vice versa.
We arrange so that the principal term, {\it i.e.}, the term with the
largest amplitude, is the one with coefficient $1-a_n(t_0)$.
Having both oscillating and nonoscillating components is almost never
an obstacle to extracting energies.  Because the two contributions are
functionally very different, there is little chance for confusion.
Because $\lambda_n(t_0, t_0) = 1$, it is useful to consider imposing
the sum rule
\begin{equation}
\Sigma_n \equiv 1 - a_n(t_0) + b_n(t_0) + c_n(t_0) + d_n(t_0) \approx 1 \,.
\label{eq:sumrule}
\end{equation}

From Eq.~(\ref{eq:eigenvalue}) we see that the parity partner energies
$E_n^\prime$ might not always be equal to the energy of a state, since
we may have either $E_n^\prime = E_m$ or $E_n^\prime = 2 E_n - E_m$,
where $E_m$ is the energy of a nearby state.  In principle the same
choices apply to the excited state values $\bar E_n$ and $\bar
E_n^\prime$, but in practice these energies could represent a weighted
average of an array of possible states including the lowest excluded
state $E_{N+1}$.

If the set of interpolating operators ${\cal O}_i$ is sufficiently
complete, we expect to be able to separate the oscillating and
nonoscillating eigenvalues, meaning that the coefficients of the
parity-partner terms should be negligible.  This implies that the
linear combination of operators
\begin{equation}
  {\cal \bar O}_n = \sum_i {\cal O}_i Z^{-1}_{i,k} \,,
\end{equation}
to a good approximation, generates a hadron correlator without an
oscillating component.  However, it often happens that the set of
operators are nearly linearly dependent.  For example, if the
interpolating operators differ only in a smearing width, we have found
that the coefficients $1-a_n$ and $c_n$ can be comparable in
magnitude.  In that case the eigenvalues contain a significant pair of
parity partners, and adding a new interpolating operator to the set
might serve, instead, to isolate an excited state, rather than a
low-lying parity partner.

\section{$D_s$ meson spectrum}

We illustrate the method by considering mesons generated by
interpolating operators consisting of a clover (Fermilab) charm quark
\cite{ElKhadra:1996mp} and a staggered strange antiquark.  The lightest of these
is the $D_s$ meson.  Previous studies of this system with variational
methods treated both quarks in the clover formalism
\cite{Mohler:2011dm,Bali:2011et,Bali:2013pp,Moir:2013mp,Mohler:2013cb}.

\subsection{Ensemble parameters}

\begin{figure}
\includegraphics{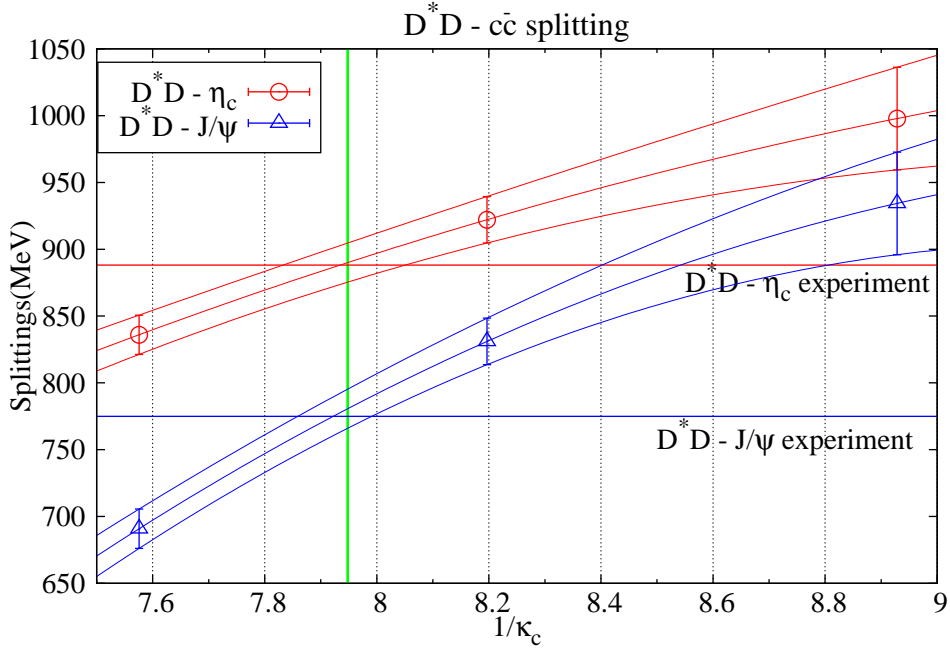}  
  \caption{The rest mass splitting $2 M(D_s) - M(\eta_c)$ as a
    function of $1/\kappa_c$, used for tuning the charm quark mass.}
 \label{fig:kappa_tune}
\end{figure}

We work with the MILC ensemble with lattice spacing $a = 0.15089(17)$
\cite{Bazavov:2014wgs} fm, generated in the presence of $2+1+1$
flavors of highly improved staggered sea quarks (HISQ), {\it i.e.},
equal up and down sea quark masses, plus strange and charm sea quarks
with all masses approximately equal to their physical values
\cite{Bazavov:2012xda}.  The lattice dimension is $32^3 \times 48$.
We measured the charm-strange meson correlator on 988 gauge
configurations separated by six molecular dynamics time units with
eight uniformly spaced source times per configuration.  The
charm-strange mesons were constructed with a clover (Fermilab) charm
quark and a strange HISQ with mass equal to the strange sea quark in
the ensemble. We also measured the charmonium correlator to set the
charm quark mass. It is tuned so that the splitting between the $D_s$
and $\eta_c$ rest masses $2 M(D_s) - M(\eta_c)$ is approximately equal
to its experimental value, as shown in Fig.~\ref{fig:kappa_tune}.  The
resulting hopping parameter is $\kappa_c =
0.1256^{+0.0021}_{-0.0014}$.

To construct the charm-strange meson we consider a variety of
single-time-slice, zero-momentum interpolating operators ${\cal O}_i$
of the form
\begin{equation}
   {\cal O}_i(t) = \sum_{\bf x}\bar Q({\bf x},t) J_i q({\bf x},t) \,,
\label{eq:interp}
\end{equation}
where $Q$ is the clover charm quark field and $q$ is the HISQ field,
converted by standard methods to a ``naive'' Dirac field according to
Eq.~(\ref{eq:naive}).  Both fields carry suppressed Dirac spin and
color indices.  The current operators $J_i$ in this study are listed
in Table~\ref{tab:interp}.  We introduce three types of covariant
Gaussian smearing, defined, as usual, in terms of the gauge-covariant
Dirac operator $\Dslash$ and a smearing width $r_x$:
\begin{equation}
     S_x = \exp(r_x^2 \Dslash^2/4)
\end{equation}
for $x = a,b,c$ with widths $r_a = 0$ (local operator), $r_b = 1.6$
(only clover quark smeared) and $r_c = \sqrt{2 r_b^2} = 2.2$ (both
clover and staggered quarks smeared).

As we have noted, the states belonging to the channels characterized
by the opposite-parity irreducible representations (irreps) $A_1^+$,
$T_1^+$ and $T_2^-$ can be extracted as the parity partners of states
in the irreps $A_1^-$, $T_1^-$ and $T_2^+$.
  
\begin{table}
  \caption{Current operators $J_i$ for constructing interpolating
    operators $\bar Q({\bf x},t) J_i q({\bf x},t)$ for the
    charm-strange mesons in this study for each of the indicated
    irreps of the octahedral group (with spatial inversions): $O_h$.
    The notation $S_x$ represents a covariant Gaussian smearing
    operator with one of three smearing widths $a, b, c$ as discussed
    in the text.  The single-time-slice operators typically generate
    states with both parities.  The indicated parity is for the
    nonoscillating state.
  \label{tab:interp}
  }
\begin{center}
    \begin{tabular}{c@{\quad}c@{\quad}c}
      \hline \hline
      $A_1^-$ & $T_1^-$ & $T_2^+$\\
      \hline
      $\gamma_5 \cdot S_{a, c}$ & $\gamma_i \cdot S_{a, b, c}$ & $\left|
      \varepsilon_{i j k} \right| \gamma_j \nabla_k$\\
      $\gamma_t \gamma_5 \cdot S_{a, c}$ & $\gamma_t \gamma_i \cdot S_{a,
      b, c}$ & $\left| \varepsilon_{i j k} \right| \gamma_t \gamma_j
      \nabla_k$\\
      $\gamma_5 \gamma_i \cdot \nabla_i$ & $\bm{I} \cdot \nabla_i$ & \\
      $\gamma_t \gamma_5 \gamma_i \cdot \nabla_i$ & $\gamma_t \cdot \nabla_i$
      & \\
      & $\varepsilon_{i j k} \gamma_5 \gamma_j \nabla_k$ & \\
      & $\varepsilon_{i j k} \gamma_t \gamma_5 \gamma_j \nabla_k$ & \\
      \hline \hline
    \end{tabular}
  \end{center}
\end{table}

\subsection{Effective energies from generalized eigenvalues}

\begin{figure}
  \includegraphics{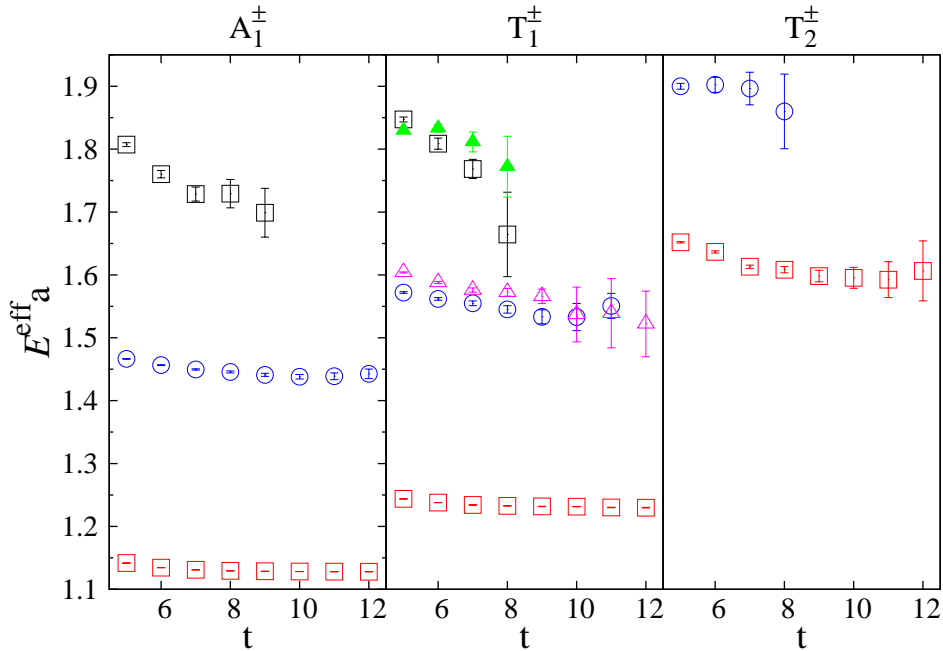}
   \caption{Smoothed effective masses ${\cal E}_{\rm{eff}}$ from the
     the eigenvalues in the $A_1^\pm$, $T_1^\pm$ and $T_2^\pm$
     charm-strange channels as a function of $t$. All interpolating operators
     listed in Table~\ref{tab:interp} are used.  The reference times are
     $t_0 = 3$ for $A_1^\pm$, $T_1^\pm$ and 
     $t_0 = 4$ for $T_2^\pm$, which are about $0.45$ fm and $0.6$ fm, 
     respectively.  The plot symbols and level assignments
     are listed in Table~\ref{tab:effmass} and discussed in
     the text.}
\label{fig:effmass}
\end{figure}

Following the procedure described in Sec.~\ref{sec:variational} we
extract the leading eigenvalues $\lambda_n(t,t_0)$ for each channel.
Then, first, we consider the corresponding effective energies.  Since
each eigenvalue could contain both oscillating (O) and nonoscillating
(NO) components, for each eigenvalue we attempt to extract effective
energies for both cases:
\begin{eqnarray}
    E^{\left( k \right)}_{\rm{eff}} \left( t \right) & = & \log \left[
    \lambda^{\left( k \right)} \left( t + 1 \right) / \lambda^{\left( k
    \right)} \left( t \right) \right] \quad \mbox{NO} \\
    E^{\left( k \right)}_{\rm{eff}} \left( t \right) & = & 
    \log \left[ - \lambda^{\left( k \right)} \left( t + 1 \right) /
    \lambda^{\left( k \right)} \left( t \right) \right] \quad \mbox{O} \,.
\end{eqnarray}
In either case, we find it helpful to smooth the result:
\begin{equation}
  {\cal E}^{\left( k \right)}_{\rm{eff}}  =  \frac{1}{4} \left[
  E_{\rm{eff}}^{\left( k \right)} \left( t + 1 \right) - 2
  E_{\rm{eff}}^{\left( k \right)} \left( t \right) +
  E_{\rm{eff}}^{\left( k \right)}(t - 1) \right] \,.
\end{equation}
We set the reference time $t_0 = 3$ (4 in the case of $T_2^+$). In the
variational calculation we include all operators in the respective
columns of Table~\ref{tab:interp}, and we examine results for all six
channels $A_1^\pm$, $T_1^\pm$, and $T_2^\pm$.  These single-time-slice
operators generate states of both parities.  The parity indicated in
the table is for the nonoscillating state.  The resulting effective
energies (masses in our zero-momentum case) for both parities are
plotted in Fig.~\ref{fig:effmass} as a function of $t$ and tabulated
in Table~\ref{tab:effmass}.

Including all interpolating operators in many cases permits a clean
isolation of the parity partners.  That is, for a given eigenvalue,
often only the oscillating or nonoscillating component is robust, and
the partner component is too weak to obtain a statistically
significant effective mass.  So only the robust states are plotted in
Fig.~\ref{fig:effmass} and listed in Table~\ref{tab:effmass}.

\begin{table}
  \caption{Classification of states identified from their effective
    masses shown in Fig.~\ref{fig:effmass}.  Listed are the eigenvalue
    indices, whether the principal state is obtained from the oscillating (O) or
    nonoscillating (NO) effective mass, the plot symbol, the inferred $O_h$
    irreducible representation and continuum spin/parity, and the
    assigned hadronic state \cite{Agashe:2014kda}, if obvious.
  \label{tab:effmass}}
  \begin{tabular}{lllll}
    \hline \hline
     $n$ & NO/O & plot symbol & $J^P$ & assignment \\
    \hline
    \multicolumn{5}{c}{$A_1^\pm$} \\
    \hline
   0 &  NO & red squares      & $A_1^-$, $0^-$ & $D_s$ \\
   1 &  O  & blue circles     & $A_1^+$, $0^+$ & $D_{s0}^\ast(2317)$ \\
   2 &  NO & black squares    & $A_1^-$, $0^-$ & ? \\
    \hline                        
    \multicolumn{5}{c}{$T_1^\pm$} \\
    \hline                        
   0 &  NO & red squares      & $T_1^-$, $1^-$ & $D_s^\ast$ \\
   1 &  O  & blue circles     & $T_1^+$, $1^+$ & $D_{s1}(2460)$ \\
   2 &  O  & purple triangles & $T_1^+$, $1^+$ & $D_{s1}(2536)$ \\
   3 &  NO & black squares    & $T_1^-$, $1^-$ & $D_{s1}^\ast(2700)$ \\
   4 &  NO & green triangles  & $T_1^-$, $1^-$ & ? \\
    \hline                        
    \multicolumn{5}{c}{$T_2^\pm$} \\
    \hline                        
   0 &  NO & red squares      & $T_2^+$, $2^+$ & $D_{s2}^\ast(2573)$ \\
   1 &  O  & blue circles     & $T_2^-$, $2^-$ & ? \\
    \hline\hline
  \end{tabular}
\end{table}

\subsection{Multiexponential fit to generalized eigenvalues}

From the foregoing effective mass analysis, we find that when all the
interpolating operators in Table~\ref{tab:interp} are in use, we
effectively isolate the low-lying parity partners.  It is interesting
to examine the progressive isolation as the dimension $N$ of the
interpolating operator basis is increased or as the reference time
$t_0$ is increased.  To do this we fit the eigenvalues to our preferred
model Eq.~(\ref{eq:fitstaggeigen}), and, for each eigenvalue, we study
the effect on the principal amplitude $1 - a_n$ and mass
$M_n$.  We discuss results for the $A_1^\pm$, $T_1^\pm$, and
$T_2^\pm$ channels separately.

For the fit range we choose $t_{\rm min} = t_0 + 1$ or $t_0 + 2$, and
we choose $t_{\rm max}$ to achieve a resonably low $\chi^2$.  For most
cases, within this fit range, two to four exponentials from our model
Eq.~(\ref{eq:fitstaggeigen}) are enough to get a robust fit result for
the chosen low $t_0$.  To impose the constraint Eq.~(\ref{eq:sumrule}),
we replace one of the amplitude parameters by $\Sigma_n$ in
Eq.~(\ref{eq:sumrule}) and constrain $\Sigma_n$ using a gaussian
Bayesian prior with central value and width ($1$, $\sigma$).  Often
the Bayesian constraint is unnecessary.

\subsubsection{$A_1^\pm$ channel}

\begin{figure}
  \begin{subfigure}
  \centering
  \includegraphics[width=13cm,height=75mm]{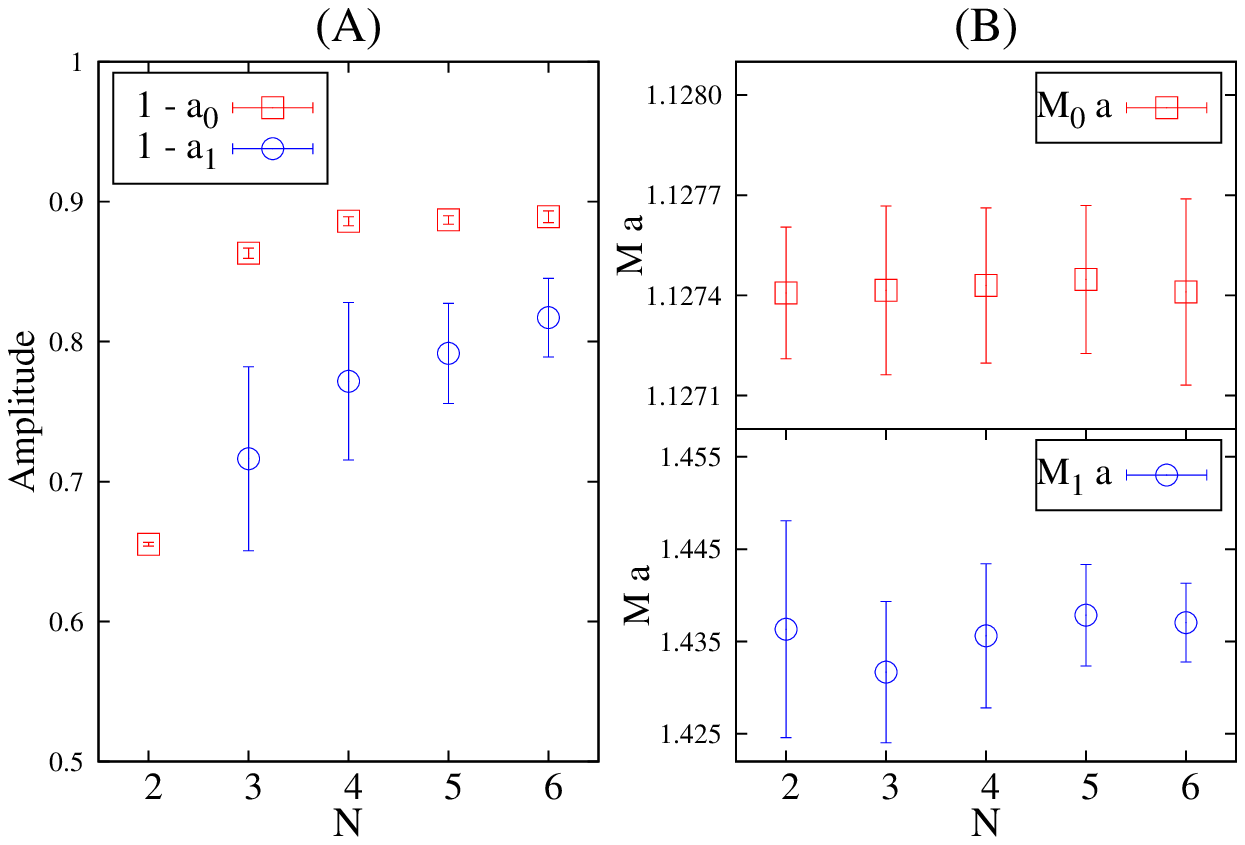} 
  \end{subfigure}
  \begin{subfigure}
  \centering
  \includegraphics[width=13cm,height=75mm]{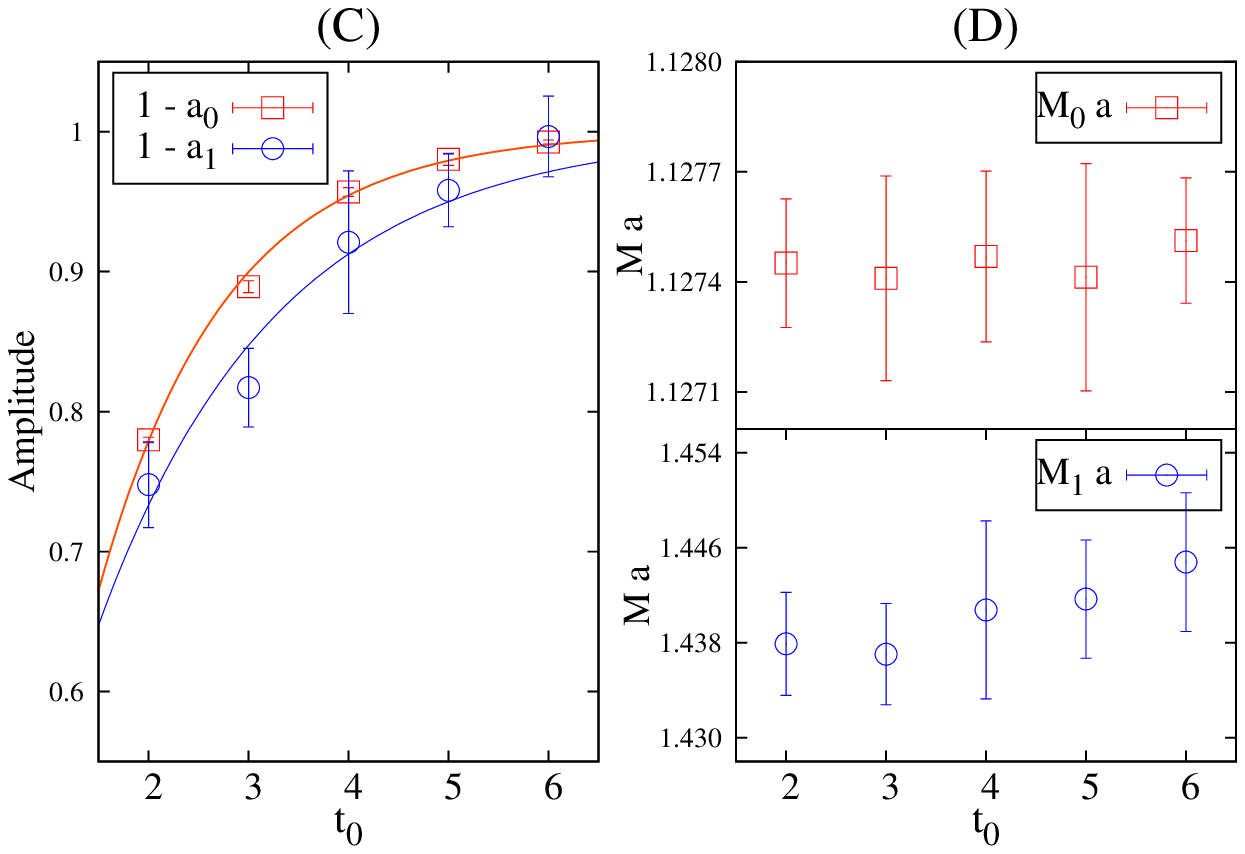}  
    \caption{Progressive improvement in the isolation of parity
      partner eigenstates with the increasing dimension $N$ of the
      interpolating operator basis.  Improvement is demonstrated for
      the two leading eigenvalues $\lambda_0$ and $\lambda_1$ in the
      $A_1^\pm$ channel by examining the principal coefficients and
      masses from a fit to Eq.~(\protect\ref{eq:fitstaggeigen}). Panels
      A and C show the principal fit coefficients $1-a_0$ and $1-a_1$
      and panels B and D, the masses $M_0 = E_0$ and $M_1 = E_1$ as a
      function of (A,C) the number of interpoling operators $N$ and
      (B,D) the reference time $t_0$. In panels A and B the
      interpolating operators are added in the order discussed in the
      text while fixing $t_0 = 3$.  In panels C and D $t_0$ is varied
      while fixing $N = 6$.  We see that in both cases the principal
      coefficient approaches one, indicating effective isolation of
      the state. The solid lines represent a fit to the function
      $1 - r_n  e^{-\Delta M_n t_0}$, adjusting both $r_n$ and $M_n$.}
\label{fig:A1amp}
  \end{subfigure}
\end{figure}

To show how results at fixed $t_0$ change as the dimension $N$ of the
interpolating operator basis is increased, we must choose a sequence
of additions to the basis.  Obviously, the result depends on how we do
that.  For the $A_1^\pm$ channel, we start at $N = 2$ with the
set $\{\gamma_5 \cdot S_{a,c}\}$ (set A).  For $N = 3$ and 4, we
include $\{\gamma_t \gamma_5 \cdot S_{a,c}\}$ (set B). Finally for $N =
5$ and 6 we include two operators involving derivatives (set C).
Results for the $A_1^\pm$ channel are shown in Fig.~\ref{fig:A1amp}
and listed in Table~\ref{tab:A1fitresult}.
Note that in Fig.~\ref{fig:A1amp}, we do not display the result for 
$n = 2$.

We find, as expected, that as $N$ increases in this way with fixed
$t_0 = 3$, the amplitude $1 - a_n$ approaches 1 and the mass
$M_n$ stabilizes.  At the same time, as shown in the table, 
the amplitude $c_0$ of the $\lambda_0$ parity
partner state decreases from $37\%$ for $N = 2$ to $0.1\%$ for $N =
6$, and the amplitude $c_1$ of the $\lambda_1$ parity partner
state decreases from $5\%$ to $0.4\%$.

For the case $N = 2$, the two ${\gamma_5 \cdot S_{a,c}}$ interpolating
operators couple in almost identical proportions to the lowest
nonoscillating and oscillating states.  So they are linearly dependent
with respect to these two states.  The result, as shown in
Table~\ref{tab:A1fitresult} set A, is that both parity partners appear
with sizeable amplitudes in the leading eigenvalue $\lambda_0$.  We
also see that adding $\{\gamma_t \gamma_5 \cdot S_{a,c}\}$ (set B) is
enough to separate the parity partners with the even parity state now
appearing in $\lambda_1$.  Finally, with the full set of operators,
the amplitudes $1 - a_n$ for the partners are greater than 0.8 in
their respective eigenvalues.  In both cases the state with the next
largest amplitude is the ``excited'' state of the same parity as the
leading state.

We note that the higher state $n = 2$ has a substantial ``excited''
state contribution $b_2$, possibly because the interpolating operators
do not have good overlap with the 2S state, and therefore they couple
strongly to other states as well.

\begin{table}
   \caption{Fit results for the eigenvalues of the $A_1^\pm$ channels
     for three different interpolating operator sets with reference
     time $t_0 = 3$.  The fit parameters $a_n$, $b_n$, $c_n$, $d_n$, $E_n$,
     $\bar E_n$, $E_n^\prime$ and $\bar E_n^\prime$ are defined in
     Eq.~(\protect\ref{eq:fitstaggeigen}). 
     In Set (A), the parity partner state is so strongly mixed 
     that $E_0^\prime$ and $E_1$ are almost degenerate. 
     In set (B) and (C), to get the reasonable fit for the ground states, 
     $4$-exponential fit is required, which is 
     $3$-nonoscillating and $1$-oscillating, instead of $2$-nonoscillating 
     and $2$-oscillating. The third nonoscillating state amplitudes and 
     masses are represented by $\tilde b_n$ and $\tilde E_n$. 
     The fit information is displayed in Table~\ref{tab:A1fitinfo}.
  \label{tab:A1fitresult} 
 }

  \begin{center}
    \begin{tabular}{l@{\quad}ll@{\quad}ll@{\quad}ll@{\quad}ll@{\quad}ll}
   \hline \hline
   \multicolumn{11}{c} { set (A): $\{\gamma_5 \cdot S_{a,c}\}$ } \\
    \hline
$n$  &  $1-a_n$  &  $E_n$  &  $b_n$  & $\bar E_n$ & $\tilde b_n$ & $\tilde E_n$
& $c_n$    &  $E_n^\prime$  & $d_n$  & $\bar E_n^\prime$  \\
    \hline

0 & $ 0.655(1)$ & $1.1274(2)$  & $0.074(6)$  &  $1.74(9)$ & $0.5(3)$ & $4.3(7)$ 
  & $-0.165(5)$ & $1.4333(49)$ & $-0.15(14)$ &  $2.56(53)$ \\

1 & $ 0.13(2)$  & $1.436(12)$  & $1.1(2)$    &  $3.42(9)$ & $0.33(7)$ & $1.95(11)$
  & $-0.017(8)$ & $1.601(79)$  & $-0.5(6)$   &  $3.0(7)$ \\
     \hline
    \end{tabular}

    \begin{tabular}{l@{\quad}ll@{\quad}ll@{\quad}ll@{\quad}ll}
		\multicolumn{9}{c} { set (B): $\{\gamma_5 \cdot S_{a,c}\}$ and 
    $\{\gamma_t \gamma_5 \cdot S_{a,c}\}$ } \\
    \hline
  $n$ &  $1-a_n$       &  $E_n$         &  $b_n$  &  $\bar E_n$   
            &  $\tilde b_n$  &  $\tilde E_n$  &  $c_n$  &  $E_n^\prime$  \\
    \hline
		0 & $0.886(4)$  & $1.1274(3)$  & $0.0142(5)$     & $1.41(28)$
                & $0.091(9)$  & $2.04(16) $  & $-0.0047(38)$ & $2.06(32)$ \\ 
		1 & $0.781(25)$ & $1.4369(43)$ & $0.151(2)$    & $1.94(12)$
                & $-$         & $-$          & $-0.049(3)$ & $1.781(26)$  \\
    \hline \hline
		\multicolumn{9}{c} { set (C): all $A_1^-$ operators in 
    Table~\ref{tab:interp} } \\
    \hline
$n$ &  $1-a_n$       &  $E_n$         &  $b_n$  &  $\bar E_n$   
            &  $\tilde b_n$  &  $\tilde E_n$  &  $c_n$  &  $E_n^\prime$  \\
    \hline
		0 & $0.889(4)$  & $1.1274(3)$   &  $0.0144(4)$    &  $1.40(26)$
                & $0.089(11)$ & $1.99(11) $   &  $-0.0010(9)$ &  $1.67(30)$  \\ 
		1 & $0.810(31)$  & $1.4361(46)$ &  $0.173(3)$     &  $1.855(87)$
                & $-$          & $-$          &  $-0.0039(18)$ &  $1.67(17) $ \\
		2 & $0.558(51)$  & $1.723(18)$  &  $0.441(6)$    &  $2.66(28)$
                & $-$          & $-$          &  $-0.081(19)$ &  $2.005(86)$ \\
    \hline
    \hline
    \end{tabular}
  \end{center}
\end{table}

\begin{table}
   \caption{Fit information for the eigenvalues of the $A_1^\pm$
     channels.  $\Sigma_n$ represents $1 - a_n + b_n + c_n + d_n$.  
     The next column shows its prior central value and width.  We
     found that it is close to $1$, as expected from the sum rule of
     Eq.~(\ref{eq:sumrule}), as long as $N$ is large enough.
  \label{tab:A1fitinfo} 
   }

  \begin{center}
    \begin{tabular}{l@{\quad}l@{\quad}l@{\quad}l@{\quad}l@{\quad}l@{\quad}l@{\quad}l}
    \hline \hline
      &  $n$ & NO/O & $\Sigma_n$ & prior $\pm$ width & fit type  & fit range &  $\chi^2/{\rm d.o.f} $ \\ 
    \hline
    \multirow{2}{*}{set (A)} 
      &  0 & NO & $1.00(2)$ & $1 \pm 0.02$ & 5-exp & $4$-$16$ & $2.0/3$ \\
      &  1 & O  & $1.00(2)$ & $1 \pm 0.02$ & 5-exp & $4$-$16$ & $2.4/3$ \\
    \hline
    \multirow{2}{*}{set (B)} 
      &  0 & NO & $0.986(4)$ & $1 \pm 0.2 $ & 4-exp & $4$-$20$ & $5.6/9$ \\
      &  1 & O  & $0.883(6)$ & $1 \pm 0.2$  & 3-exp & $4$-$18$ & $6.8/9$ \\
    \hline
    \multirow{3}{*}{set (C)} 
      & 0 & NO  & $0.992(3)$  & $1 \pm 0.1$  & 4-exp & $4$-$20$ & $5.5/9$ \\
      & 1 & O   & $0.978(5)$  & $1 \pm 0.1$  & 3-exp & $4$-$18$ & $7.3/9$ \\
      & 2 & NO  & $0.944(87)$ & $1 \pm 0.1$  & 3-exp & $4$-$11$ & $3.3/2$ \\
    \hline \hline
    \end{tabular}
  \end{center}
\end{table}
                             
Instead of varying $N$ at fixed $t_0$ we can vary $t_0$ at fixed $N$.
We find that as $t_0$ increases with fixed $N = 6$, the amplitude $1 -
a_n$ also approaches 1, and the mass $M_n$ stabilizes.  We can be
slightly more quantitative here.  From
Eqs.~(\ref{eq:eigenvalue}),(\ref{eq:eigenvalueamp01}) and
(\ref{eq:eigenvalueamp02}), we see that the coefficient $a_n$ in
Eq.~(\ref{eq:fitstaggeigen}) all tend to decrease exponentially with
$t_0$ at fixed $N$ as
\begin{equation}
   e^{-(E_{N+1}-E_n) t_0} A_{n,n,N+1} \,,
\label{eq:mainstateamp}
\end{equation}
whereas the coefficients $b_n$, $c_n$, and $d_n$ decrease
exponentially according to
\begin{equation}
e^{-2(E_{N+1}-E_n) t_0} A_{n,m,N+1}^2 \, .
\label{eq:prtstateamp}
\end{equation}
We note that at fixed $N$, the coefficient $A_{n,m,N+1}^2$ is
constant.  Indeed, as shown in Fig.~\ref{fig:A1amp} panel C, the
coefficient $1 - a_n$ can be fit with the exponential form $1 - r_n
\exp(-\Delta M_n t_0)$, where $r_n$ and $\Delta M_n$ are adjusted
to their best fit values.

We also note that the mass values $M_0$ and $M_1$ for $t_0 \ge 2$ are
statistically consistent, which justifies using low reference times
$t_0$ in conjunction with a multiexponential fit, such as
Eq.~(\ref{eq:fitstaggeigen}), to compensate for unsuppressed
contributions from other states.  The $t_0 = 6$ value (about $0.9$ fm)
was obtained from a single exponential fit, because the data were then
insufficient to determine excited state contributions.

\begin{figure}
  \begin{subfigure}
  \centering
  \includegraphics[width=13cm,height=8cm]{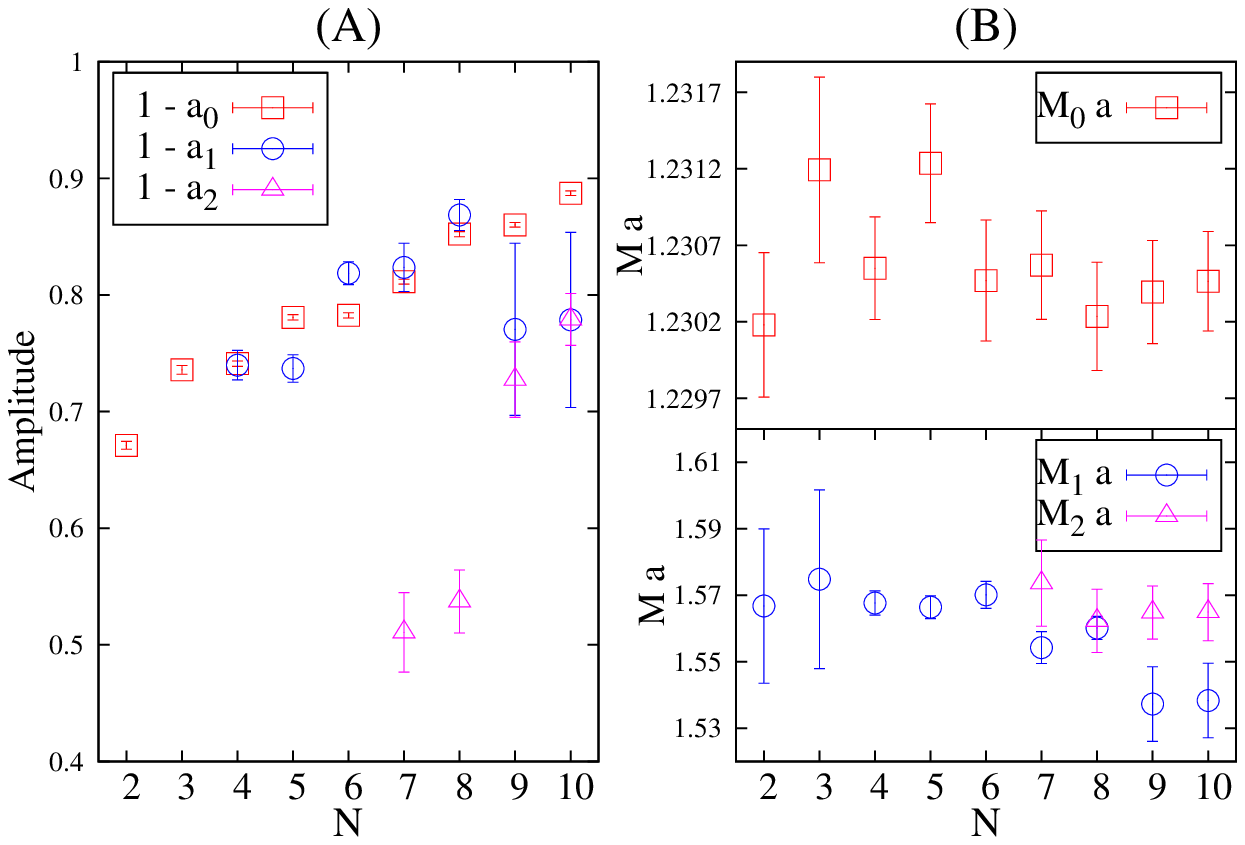}
  \end{subfigure}
  \begin{subfigure}
  \centering
  \includegraphics[width=13cm,height=8cm]{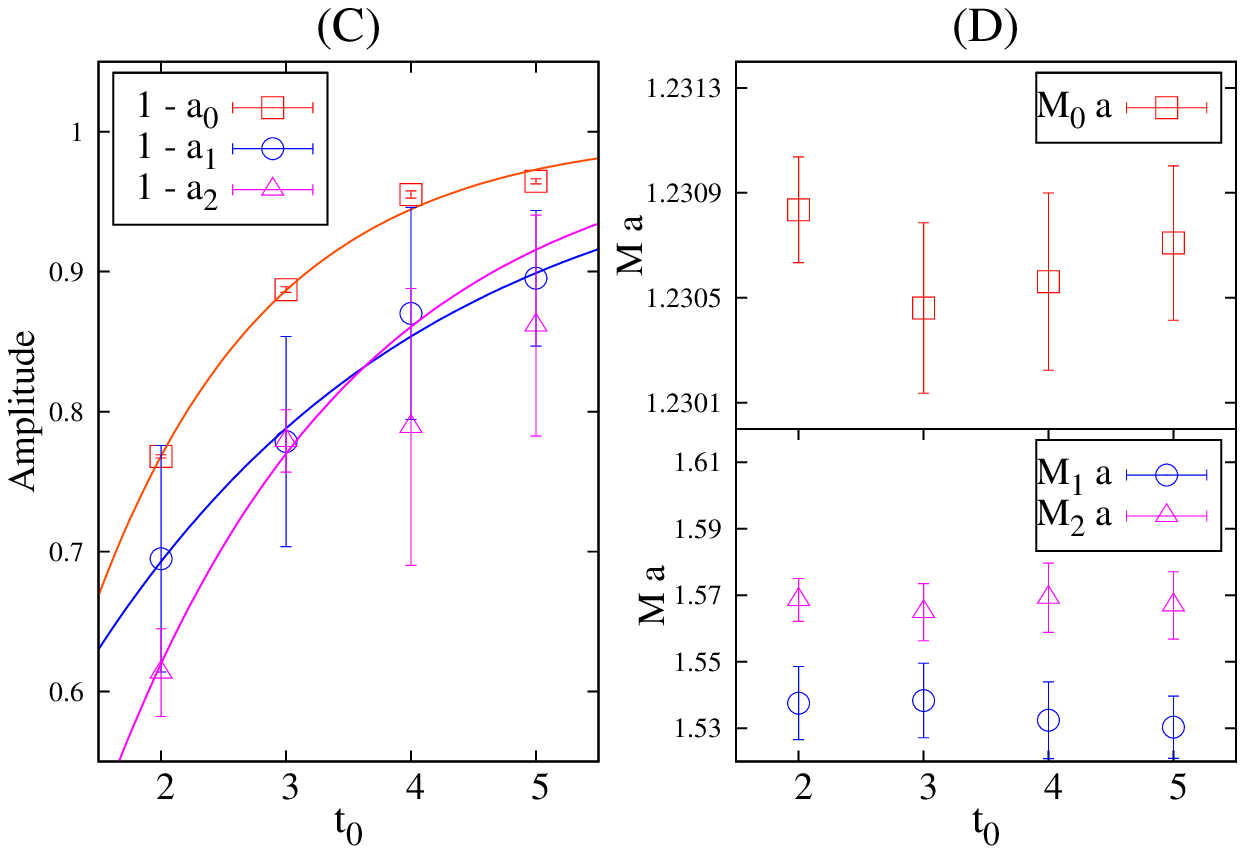}
    \caption{ As in Fig.~\ref{fig:A1amp}, but for the lowest two
      states in the $T_1^\pm$ channels.
      The operator sets are defined in the text.}
  \label{fig:T1amp}
  \end{subfigure}
\end{figure}
%

%
\begin{table}
	\caption{Fit results for the eigenvalues of the  $T_1^\pm$ channels 
          for three different interpolating operator sets with reference
          time $t_0 = 3$.  The notation is the same as in
          Table~\protect\ref{tab:A1fitresult}.  \label{tab:T1fitresult}
}
  \begin{center}
   \begin{tabular}{l@{\quad}ll@{\quad}ll@{\quad}ll@{\quad}ll@{\quad}ll}
   \hline \hline
   \multicolumn{11}{c} { set (A): $\{\gamma_5 \cdot S_{a,c}\}$ } \\
    \hline
$n$  &  $1-a_n$  &  $E_n$  &  $b_n$  & $\bar E_n$ & $\tilde b_n$ & $\tilde E_n$
& $c_n$    &  $E_n^\prime$  & $d_n$  & $\bar E_n^\prime$  \\
    \hline

0 & $0.671(3)$  & $1.2302(5)$ & $0.071(1)$ &  $1.76(13)$ & $0.6(3)$ & $3.83(29)$ 
  & $-0.142(6)$ & $1.523(7)$  & $-0.2(3)$  &  $2.85(77)$ \\

1 & $0.051(9)$ & $1.567(23)$  & $0.41(4)$ &  $2.31(13)$ & $1.0(3)$ & $4.9(1.7)$
  & $-0.06(2)$ &  $1.69(5)$  & $-0.4(1)$  &  $2.4(2)$ \\
    \hline
    \end{tabular}

    \begin{tabular}{l@{\quad}ll@{\quad}ll@{\quad}ll@{\quad}ll}
		\multicolumn{9}{c} { set (B): $\{\gamma_i \cdot S_{a,b,c}\}$ and
    $\{\gamma_t \gamma_i \cdot S_{a,b,c}\}$ } \\
    \hline
$n$    & $1-a_n$  &  $E_n$         & $b_n$  & $\bar E_n$   
    & $c_n$    &  $E_n^\prime$  & $d_n$  & $\bar E_n^\prime$  \\
    \hline
		0 &  $0.784(2)$  &  $1.2306(4)$   & $0.164(3)$  &  $2.179(99)$  
                &  $-0.073(2)$ &  $1.5382(76)$  & $-$        &  $-$ \\
		1 &  $0.798(33)$  &  $1.5647(92)$  & $0.286(2)$   &  $2.94(87)$ 
                &  $-0.081(20)$ &  $1.667(71)$   & $-$         &  $-$ \\
		2 &  $0.276(49)$  &  $1.669(49)$  & $0.819(1)$   &  $2.44(14)$ 
                &  $-0.097(10)$ &  $1.648(30)$  & $-$         &  $-$ \\
    \hline
		\multicolumn{9}{c} { set (C): all $T_1^-$ operators listed  
    in Table~\ref{tab:interp} }\\
    \hline
$n$ & $1-a_n$  &  $E_n$         & $b_n$  & $\bar E_n$   
      & $c_n$    &  $E_n^\prime$  & $d_n$  & $\bar E_n^\prime$  \\
   \hline
 		0 &  $0.872(2)$   &  $1.2305(3)$  & $0.0856(3)$  &  $1.879(35)$  
                &  $-0.0108(8)$ &  $1.586(28)$  & $-$        &  $-$ \\
		1 &  $0.774(80)$  &  $1.538(12)$  & $0.187(6)$  &  $1.94(19)$ 
                &  $-0.012(3)$  &  $1.59(10)$   & $-$        &  $-$ \\
		2 &  $0.761(47)$   &  $1.569(10)$ & $0.256(17)$   &  $2.44(66)$ 
                &  $-0.0020(29)$ &  $1.27(25)$  & $-0.06(19)$ &  $2.6(1.8)$ \\
		3 &  $0.529(90)$  &  $1.739(31)$ & $0.482(4)$   &  $2.34(12)$ 
                &  $-0.017(3)$  &  $1.556(68)$   & $-$         &  $-$ \\
		4 &  $0.720(23)$  &  $1.8248(89)$  & $0.504(3)$   &  $4.11(55)$ 
                &  $-0.22(5)$   &  $2.323(91)$   & $-$         &  $-$ \\
    \hline  \hline
    \end{tabular}
  \end{center}
\end{table}

\begin{table}
   \caption{Fit information for the eigenvalues of the $T_1^\pm$ channels.
   The notation is the same as in Table~\ref{tab:A1fitinfo}.   \label{tab:T1fitinfo} 
}

  \begin{center}
    \begin{tabular}{l@{\quad}l@{\quad}l@{\quad}l@{\quad}l@{\quad}l@{\quad}l@{\quad}l}
    \hline \hline
    & $n$ & NO/O & $\Sigma_n$ & prior $\pm$ width & fit type  & fit range &  $\chi^2/{\rm d.o.f} $ \\ 
    \hline
    \multirow{2}{*}{set (A)} 
     & 0 & NO & $1.00(5)$ & $1 \pm 0.05$ & 5-exp & $4$-$17$ & $5.9/4$ \\
     & 1 & O  & $1.00(5)$ & $1 \pm 0.05$ & 5-exp & $4$-$17$ & $3.5/4$ \\
    \hline
    \multirow{2}{*}{set (B)} 
     & 0 & NO & $0.87(3)$  & $1 \pm 0.3$  & 3-exp & $5$-$19$ & $6.5/9$ \\
     & 1 & O  & $1.00(30)$ & $1 \pm 0.3$  & 3-exp & $5$-$17$ & $5.9/7$ \\
     & 2 & NO & $0.97(9)$ & $1 \pm 0.1$  & 3-exp & $5$-$11$ & $0.4/1$ \\
    \hline
    \multirow{3}{*}{set (C)} 
     & 0 & NO & $0.962(2)$  & $1 \pm 0.1$ & 3-exp & $4$-$20$ & $11.2/11$\\
     & 1 & O  & $0.952(11)$ & $1 \pm 0.1$ & 3-exp & $4$-$17$ & $5.2/8$ \\
     & 2 & O  & $0.948(58)$ & $1 \pm 0.1$ & 4-exp & $4$-$13$ & $0.9/2$ \\
     & 3 & NO & $0.994(19)$ & $1 \pm 0.02$ & 3-exp & $4$-$10$ & $2.2/1$ \\
     & 4 & NO & $0.991(99)$ & $1 \pm 0.1$  & 3-exp & $4$-$11$ & $1.3/2$ \\
    \hline \hline
    \end{tabular}
  \end{center}
\end{table}

\subsubsection{$T_1^\pm$ channel}

In Fig.~\ref{fig:T1amp} and Table~\ref{tab:T1fitresult}, we show
the progressive isolation of low-lying parity partners in the
$T_1^\pm$ channel.  For $N = 2$ we use only $\{\gamma_i \cdot
S_{a,c}\}$ (set A).  For $N = 3$, 4, 5, and 6, we include $\{\gamma_t
\gamma_i \cdot S_{a,b,c}\}$ (set B), respectively.  Finally, we
include the remaining operators in Table~\ref{tab:interp} involving
derivatives to reach $N = 10$ (set C).
Note that in Fig.~\ref{fig:T1amp}, we do not show the results for 
$n = 3$ and $4$. 

As with the $A_1^\pm$ case we find that a set of operators that differ
only by their degree of smearing (set A) is ineffective in separating
the parity partners, so eigenvalue $\lambda_0$ contains both of them.
However, unlike the $A_1^\pm$ channel, the $T_1^\pm$ channel has two
fairly closely spaced $T_1^+$ states.  So the oscillating term in
$\lambda_0$ in set A could represent a mixture of both.  Adding the
$\{\gamma_t \gamma_i \cdot S_{a,b,c}\}$ operators helps partly in
separating the states, but $\lambda_0$ for that set still has a strong
oscillating component.  A nearly complete separation occurs only after
nine or more operators are included (set C).  Then $\lambda_0$
contains only the nonoscillating state and the two oscillating states
appear separately in $\lambda_1$ and $\lambda_2$.

We also see that for higher excitations, the separation of states is
less clean.  Level $n=3$ requires a substantial amplitue $b_3$,
and level $4$ requires a substantial opposite-parity amplitude
$c_4$.

Panels C and D in Fig.~\ref{fig:T1amp} show the principal-state
amplitudes and masses from fits as a function of the reference time,
$t_0$.  Since $N = 10$ is fixed, $A_{n,n,N+1}$ is constant, and the
amplitude $a_0$ should decrease exponentially according to
Eq.~(\ref{eq:mainstateamp}).  When $t_0 = 5$, the opposite parity
contributions become negligible, and the mass can be extracted to good
approximation with a single-exponential fit.  Even so, as shown in
panel D, all masses are statistically equivalent over the whole range
of $t_0$ displayed.  This equivalence suggests that one can extract
the desired mass for low $t_0$ using a multi-exponential fit with
three or four exponentials, as in Eq.~(\ref{eq:staggmultiexp}).


\subsubsection{$T_2^\pm$ channel}

\begin{table}
  \caption{ Fit results for the eigenvalues of the $T_2^\pm$ channels
    with reference time $t_0 = 4$. The notation is the same as 
    in Table~\protect\ref{tab:A1fitresult}.
  \label{tab:T2fitresult}
 }
  \begin{center}
    \begin{tabular}{l@{\quad}ll@{\quad}ll@{\quad}ll@{\quad}ll}
     \hline \hline
		\multicolumn{9}{c} { set (C): all $T_2^+$ operators listed  
    in Table~\ref{tab:interp} }\\
    \hline
$n$ & $1-a_n$  &  $E_n$         & $b_n$  & $\bar E_n$   
    & $c_n$    &  $E_n^\prime$  & $d_n$  & $\bar E_n^\prime$  \\
    \hline
		0 & $1.01(11)$  & $1.594(16)$ & $0.338(6)$ & $2.17(27)$ 
                & $-0.34(15)$ & $1.890(18)$ & $-$      & $-$ \\
		1 & $1.34(3)$   & $1.903(12)$ & $-$      & $-$ 
                & $-0.35(4)$  & $2.11(7)$   & $-$      & $-$ \\
    \hline \hline
    \end{tabular}
  \end{center}
\end{table}
\begin{table}
   \caption{Fit information for the eigenvalues of the $T_2^\pm$ channels.
   The notation is the same as in Table~\ref{tab:A1fitinfo}.
  \label{tab:T2fitinfo} 
}

  \begin{center}
    \begin{tabular}{l@{\quad}l@{\quad}l@{\quad}l@{\quad}l@{\quad}l@{\quad}l@{\quad}l}
    \hline \hline
    & $n$ & NO/O & $\Sigma_n$ & prior $\pm$ width & fit type  & fit range &  $\chi^2/{\rm d.o.f} $ \\ 
    \hline
    \multirow{2}{*}{set (A)} 
     & 0 & NO & $1.009(60)$ & $1 \pm 0.08$ & 3-exp & $5$-$15$ & $4.1/5$ \\
     & 1 & O  & $0.99(1)$ & $-$ & 2-exp & $5$-$13$ & $4.5/5$ \\
    \hline \hline
    \end{tabular}
  \end{center}
\end{table}

Finally, Table~\ref{tab:T2fitresult} lists fit results for the
$T_2^\pm$ channel.  Because there are only a few interpolating
operators, the parity partners are not well separated at $t_0 = 4$.
Thus even at reasonably low $t_0$, the multiexponential fit again
helps to compensate for contamination from other unsuppressed
exponential contributions.

\subsection{Comparison with observed states}

\begin{figure}
\includegraphics{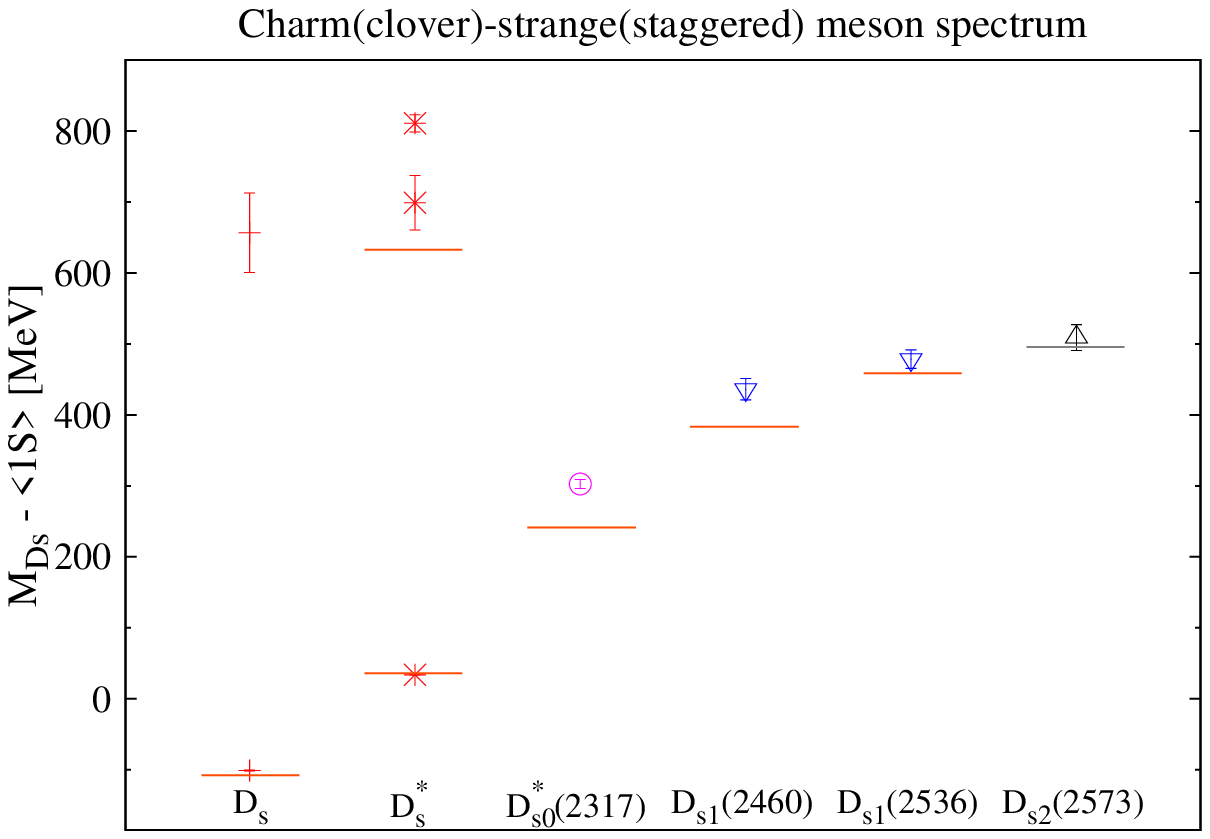}  
  \caption{Comparison of our crude theoretical charm-antistrange meson
    spectrum (symbols with errors) with experimental values (short
    horizontal lines) with tentative assignments of the levels.  Mass
    splittings are shown relative to the spin-averaged $D_s$ $1S$ state,
    namely $\overbar{1S} = \frac{1}{4}(D_s + 3 D^\ast_s)$. }
  \label{fig:Ds_spectrum}  
\end{figure}

Even though we are working at only one lattice spacing with quark
masses close, but not finely tuned, to their physical values, and we
have not considered effects of two-meson channels, it is tempting to
compare our results with the experimentally known
masses\cite{Agashe:2014kda}.  This is done in
Fig.~\ref{fig:Ds_spectrum} and Table~\ref{tab:splittings}, including
tentative assignments.
\begin{table}
   \caption{Mass splittings in the $D_s$ spectrum. 
   The experimental splittings are calculated relative to the spin-averaged
   $D_s$ $1S$ state, based on values in Ref.~\cite{Agashe:2014kda}.
  \label{tab:splittings} 
}

  \begin{center}
    \begin{tabular}{c@{\quad}c@{\quad}c}
    \hline \hline
                  & Experiment [MeV]  & Lattice [MeV]   \\
    \hline
    \hline
    $D_s^\ast - D_s$  & $143.8 \pm 0.4$   & $134.77 \pm 0.51$  \\
    \hline
    $D_s - \overbar{1S} $        &  $-107.9 \pm 0.5$ & $-101.08 \pm 0.38$  \\
    \hline
    $D_s^\ast - \overbar{1S}$    &  $35.9 \pm 0.6$   & $33.69 \pm 0.13$  \\
    \hline
    $D_{s1}^\ast(2700) - \overbar{1S}$ & $632.7 \pm 4$ & $698.9 \pm 38.4$ \\
    \hline
    $D_{s0}(2317) - \overbar{1S}$ &  $241.5 \pm 0.7$ & $302.6 \pm 6.3$ \\
    \hline
    $D_{s1}(2460) - \overbar{1S}$ &  $383.3 \pm 0.7$ & $436.2 \pm 15.0$ \\
    \hline
    $D_{s1}(2536) - \overbar{1S}$ &  $458.8 \pm 0.4$ & $478.5 \pm 13.1$ \\
    \hline
    $D_{s2}(2573) - \overbar{1S}$ &  $496.3 \pm 1.0$ & $508.9 \pm 18.3$ \\
    \hline
    \hline
    \end{tabular}
  \end{center}
\end{table}

\section{Conclusion}

The variational method is widely used to determine the eigenenergies
of the lattice QCD hamiltonian.  With this method the variational
basis is constructed by acting on the vacuum with a linear combination
of a variety of interpolating operators of appropriate conserved
quantum numbers.  The eigenvalues of a resulting generalized
eigenvalue problem then determine the eigenenergies.  We described an
extension of the method to single-time-slice interpolating operators
involving staggered fermions where the effective transfer matrix has
both negative and positive eigenvalues.

We presented a straightforward generalization of the perturbative
treatment of the ALPHA Collaboration \cite{Blossier:2009kd} that
provides an estimate of the error in the variational eigenvalue
estimates resulting from the truncation to a finite interpolating
operator basis.  Motivated by the perturbative treatment, we presented
a simple multiexponential expansion of the eigenvalues for a more
accurate determination of the energy levels.  The multiexponential
approach allows one to relax, to some extent, impractical constraints
that require large reference times in the generalized eigenvalue
problem.

We illustrated the method with a lattice QCD study of the orbital and
radial excitations of the $D_s$ meson.  In this calculation, the charm
quark was modeled in the clover fermion formulation (Fermilab
interpretation) and the strange quark, in the highly improved
staggered quark (HISQ) formulation.  All quarks, including sea quarks,
had approximately physical masses.  We found that with a sufficiently
large and diverse basis, the variational method is capable of
separating low-lying parity-partner states, placing them in separate
eigenvalues of the transfer matrix.  We showed that large reference
times lead to the suppression of extraneous multiexponential
contributions, as expected.  Finally, we compared our results for the
excitations with the experimental values and found satisfactory
agreement, considering the coarseness of the lattice and the omission
of multihadron interpolating operators.

\section*{Acknowledgment}

We thank Daniel Mohler and Heechang Na for useful discussions.  This
work was supported by the U.S. National Science Foundation under grant
NSF PHY10-034278 and the U.S. Department of Energy under grant
DE-FC02-12ER41879.  Computations were carried out on the LQCD clusters
at Fermilab and at the Utah Center for High Performance Computing.

\bibliographystyle{apsrev4-1}
\bibliography{variational_staggered}

\end{document}